\documentclass[preprint,aps,amsmath,showpacs,floatfix]{revtex4}
\newcommand{\rd}{{\rm d}}
\newcommand{\re}{{\rm e}}
\newcommand{\ri}{i}
\newcommand{\refg}[1]{(\ref{#1})}
\usepackage{graphicx}

\begin{document}

\title{A New Form of Path Integral for the
Coherent States Representation and its Semiclassical Limit}

\author{L.C. dos Santos and M.A.M. de Aguiar }

\affiliation{Instituto de F\'{\i}sica `Gleb Wataghin'\\
Universidade Estadual de Campinas, Unicamp\\
13083-970, Campinas, S\~{a}o Paulo, Brasil\\}

\begin{abstract}

The overcompleteness of the coherent states basis leads to a
multiplicity of representations of Feynman's path integral. These
different representations, although equivalent quantum
mechanically, lead to different semiclassical limits. Two such
semiclassical formulas were derived in \cite{Bar01} for the two
corresponding path integral forms suggested by Klauder and
Skagerstan in \cite{Klau85}. Each of these formulas involve
trajectories governed by a different classical representation of
the Hamiltonian operator: the P representation in one case and the
Q representation in other. In this paper we construct a third
representation of the path integral whose semiclassical limit
involves directly the Weyl representation of the Hamiltonian
operator, i.e., the classical Hamiltonian itself.

\end{abstract}


%03.65.Db Functional analytical methods
%03.65.Sq Semiclassical theories and applications

\maketitle

%%%%%%%%%%%%%%%%%%%%%%%%%%%%%%%%%%%%%%%%%%%%%%%%%%%%%%%%
%%%%%%%%%%%%%%%%%%%%%%%%%%%%%%%%%%%%%%%%%%%%%%%%%%%%%%%%
\section{Introduction}

In recent years there has been a renewed interest in semiclassical
approximations with coherent states. These approximations have
found applications in many areas of physics and chemistry. The
semiclassical coherent state propagator has a long history, that
starts with Klauder \cite{Klau78,Klau79,Klau87a} and Weissman
\cite{Weis82b}. Several properties of the propagator were
subsequently studied for a number of fundamental quantum processes
(see, e.g.,
\cite{Ada89,Shu95,Xavi,Gros98b,98ivr,Heller02,Heller03,Pol03,Par03,Rib04,Piz04}).
More recently, a detailed derivation of the semiclassical
propagator for systems with one degree of freedom was presented in
\cite{Bar01}.

The set of coherent states forms an non-orthogonal over-complete
basis, since each state in the set can be written as a linear
combination of the others. This overcompleteness, on the other
hand, has important consequences in the path integral formulation
of the propagator. It implies in the existence of several forms of
the path integrals, all equivalent quantum mechanically, but each
leading to a slightly different semiclassical limit. Klauder and
Skagerstam \cite{Klau85} proposed two basic forms for the coherent
state path integral, each of them having their corresponding
advantages and problems \cite{Klau85}. The semiclassical limit of
these two basic propagators were considered in \cite{Bar01} where
it was shown that both propagators can written in terms of
classical complex trajectories, each governed by different
classical representation of the Hamiltonian operator $\hat{H}$:
the P representation in one case and the Q representation in
other. We briefly review these representations in section 2. The
phase appearing in these semiclassical formulas turns out to be
not just the action of the corresponding complex classical
trajectory, but it also contains a `correction term' $I$ that
comes with different signs in each formula (see Eqs.(\ref{glg119})
and (\ref{glg119a})).

In \cite{Bar01} it was also suggested that a semiclassical
representation involving directly the Weyl representation of
$\hat{H}$, or the {\it classical Hamiltonian}, could probably be
constructed, and a formula for this representation was
conjectured. In this paper we derive this formula by constructing
a new representation of the quantum mechanical path integral,
Eq.(\ref{5mb8}), and deriving its semiclassical limit
Eq.(\ref{new6}). We show that the classical trajectories involved
in this formula are governed by the average between the P and Q
representations of the Hamiltonian operator. The correction term
in the phase, on the other hand, turns out to be one half of the
difference between the corresponding terms in the previous
formulations. We then show that this average Hamiltonian can be
replaced by the classical one and the correction term discarded,
the error being of order $\hbar^2$. Our final result is the
conjectured formula, Eq.(\ref{coe0}).

This paper is organized as follows: in section 2 we review the
path integral constructions of Klauder and Skagerstan
\cite{Klau85} and their semiclassical approximations \cite{Bar01}.
In section 3 we construct the new quantum representation and
derive its semiclassical limit. Finally in section 4 we show that,
within the validity of the approximations, this formula coincides
with the equation suggested in \cite{Bar01}.

%%%%%%%%%%%%%%%%%%%%%%%%%%%%%%%%%%%%%%%%%%%%%%%%%%%%%%%%
%%%%%%%%%%%%%%%%%%%%%%%%%%%%%%%%%%%%%%%%%%%%%%%%%%%%%%%%
\section{The coherent state propagator and its semiclassical approximations}

%%%%%%%%%%%%%%%%%%%%%%%%%%%%%%%%%%%%%%%%%%%%%%%%%%%%%%%%%%%%%%%%%%%
\subsection{The propagator}
The coherent state $|z\rangle$ of a harmonic oscillator of mass $m$ and
frequency $\omega$ is defined by
\begin{equation}
  \label{glg48}
  |z\rangle = \re^{-\frac{1}{2}|z|^2}\re^{z\hat{a}^\dagger}|0\rangle
\end{equation}
with $|0\rangle$ the harmonic oscillator ground state and
\begin{equation}
  \label{glg49}
  \hat{a}^\dagger = \frac{1}{\sqrt{2}}\left( \frac{\hat{q}}{b}-\ri
              \,\frac{\hat{p}}{c} \right), \qquad
  z =  \frac{1}{\sqrt{2}}\left( \frac{q}{b}+\ri
              \,\frac{p}{c} \right).
\end{equation}
In the above $\hat q$, $\hat p$, and $\hat{a}^\dagger$ are
operators; $q$ and $p$ are real numbers; $z$ is complex.  The
parameters $b = {(\hbar/ m \omega )}^{\frac{1}{2}}$ and $c =
{(\hbar m \omega )}^{\frac{1}{2}}$ define the length and momentum
scales, respectively, and their product is $\hbar$.

For a time-independent Hamiltonian operator $\hat{H}$, the propagator
in the coherent states representation is the matrix element of the
evolution operator between the states $|z^\prime \rangle$ and $| z''
\rangle$:
\begin{equation}
  \label{glg51}
  K(z'',T;z',0) = \langle z'' | \re^{-\frac{\ri }{\hbar}\hat{H}T}
                  | z^\prime \rangle.
\end{equation}
We restrict ourselves to Hamiltonians that can be expanded in a
power series of  the creation and annihilator operators
$\hat{a}^\dagger$ and $\hat{a}$.

In the derivation the semiclassical limit of the propagator, the
Hamiltonian operator $\hat{H}$ is somehow replaced by a classical
Hamiltonian function $H(q,p)$. This `replacement', however, is not
uniquely defined, and the ambiguities that exist in the relation
between the operator $\hat{H}$ and the function $H(q,p)$ also
arise in connection with the overcompleteness of the coherent
state basis, as we shall see in the next subsections.

There are actually many ways to associate a classical function of
position and momentum $A(q,p)$ to a quantum mechanical operator
$\hat{A}$. However, three of them are specially important. The
first one, denoted $A_Q(q,p)$, is called the Q representation of
the operator $\hat{A}$ and is constructed as follows: one writes
$\hat{A}$ in terms of the creation and annihilation operators
$\hat{a}^\dagger$ and $\hat{a}$ in such a way that all the
creation operators appear to the left of the annihilation
operators, making each monomial of $\hat{A}$ look like
$c_{nm}\hat{a}^{\dagger n} {\hat{a}}^m$. Then we replace $\hat{a}$
by $z$ and $\hat{a}^\dagger$ by $z^\star$. The inverse of this
operation, that associates a quantum operator to a classical
function, is called `normal ordering'. In this case one first
writes the classical function in terms of $z$ and $z^\star$, with
all the $z^\star\,$'s to the left of the $z$'s, and then replace
$z$ by $\hat{a}$ and $z^\star$ by $\hat{a}^\dagger$.

The second possibility, called the P representation of $\hat{A}$,
is obtained by a similar procedure, but this time the monomials of
$\hat{A}$ are written in the opposite order, such that they look
like $c_{nm}\hat{a}^n \hat{a}^{\dagger m}$. Once the operator has
been put in this form one replaces again $\hat{a}$ by $z$ and
$\hat{a}^\dagger$ by $z^\star$ to obtain $A_P(q,p)$. The inverse
of this operation is called `anti-normal ordering'. Notice that
the differences between the two representations come from the
commutator of $\hat{q}$ and $\hat{p}$, which is proportional to
$\hbar$. Therefore, these differences go to zero as $\hbar$ goes
to zero.

There is, finally, a third representation which is the most
symmetric of all, and therefore the most natural. It is given by
the Wigner transformation
\begin{equation}
  \label{wig1}
A_W(q,p) = \int\rd s\,\re^{\frac{\ri}{\hbar}ps}\left\langle
q-\frac{s}{2}\left|
   \hat{A}\right|q+\frac{s}{2}\right\rangle \;.
\end{equation}
$A_W(q,p)$ is called the Weyl representation of $\hat{A}$
\cite{Hill84,Alf98}. Its inverse transformation consists in
writing the classical function in terms of $z$ and $z^\star$
considering all possible orderings for each monomial and making a
symmetric average between all possibilities before replacing $z$
and $z^\star$ by the corresponding operators. As an illustration
of these three representations we take
\begin{displaymath}
\hat{H} = \frac{1}{2} \frac{\partial^2}{\partial x^2} +
\frac{1}{2} x^2 + x^4
\end{displaymath}
($m=\hbar=1$) for which we obtain
\begin{displaymath}
\begin{array}{l}
H_Q = \frac{1}{2} (p^2 + x^2) + x^4 + \frac{1}{4}(b^2+b^{-2}) + 3b^2x^2 + 3b^4/4 \\
H_P = \frac{1}{2} (p^2 + x^2) + x^4 - \frac{1}{4}(b^2+b^{-2}) - 3b^2x^2 + 3b^4/4 \\
H_W = \frac{1}{2} (p^2 + x^2) + x^4
\end{array}
\end{displaymath}
where $b$ is the width of the coherent state. Notice the term
proportional to $x^2$ that appears with opposite signs in $H_Q$
and $H_P$, really modifying the classical dynamics with respect to
$H_W$.

In the next subsections we shall see how these different
representations appear naturally in the semiclassical limit of the
coherent state propagator.

%%%%%%%%%%%%%%%%%%%%%%%%%%%%%%%%%%%%%%%%%%%%%%%%%%%%%%%%%%%%%%%%%%%
\subsection{Basic Path Integrals and their Semiclassical Approximations}

The calculation of the semiclassical propagator in the coherent
state representation starting from path integrals was discussed in
detail in \cite{Bar01}. In this section we summarize these
previous results emphasizing the non-uniqueness of the
semiclassical limit as a consequence of the overcompleteness of
the coherent state representation. The reader is referred to
\cite{Bar01} for the details.

In order to write a path integral for $ K(z'',T;z',0)$, the time
interval has to be divided into a large number of slices and, for
each slice, an infinitesimal propagator has to be calculated. As
pointed out by Klauder and Skagerstam \cite{Klau85}, there are at
least two different ways to do that. Each of these gives rise to a
different representation of the path integral. Although they
correspond to identical quantum mechanical quantities, their
semiclassical approximations are different. We review the
construction of these two representations below.

The first form of path integral is constructed by breaking the time
interval $T$ into $N$ parts of size $\tau$ and inserting the unit
operator
\begin{equation}
  \label{glg21}
  \openone = \int|z\rangle\frac{\rm{d}^2z}{\pi}\langle z|
\end{equation}
everywhere between adjacent propagation steps. We denote the real and
imaginary parts of $z$ by $x$ and $y$, respectively. In all
integrations, ${\rm d}^2z/\pi $ means $\rd x\rd y/\pi$.  After the
insertions, the propagator becomes a $2(N-1)$--fold integral over the
whole phase space
\begin{equation}
  \label{glg82}
  K(z'',t;z',0) = \int \Bigl\{ \prod_{j=1}^{N-1}\frac{\rd^2z_j}{\pi}
                   \Bigr\} \prod_{j=0}^{N-1} \Bigl\{ \langle z_{j+1} |
                   \re^{-\frac{\ri }{\hbar}
                   \hat{H}(t_j)\tau} | z_j \rangle
                   \Bigr\}
\end{equation}
with $z_N = z''$ and $z_0=z'$. Using the coherent state overlap
formula
\begin{equation}
  \label{glg50}
 \langle z_{j+1} | z_j \rangle
                        = \exp\left\{ -\frac{1}{2}|z_{j+1}|^2
                        + z_{j+1}^\star z_j
                        - \frac{1}{2}|z_j|^2\right\}
\end{equation}
and expanding $e^{-iH\tau/\hbar} \approx 1 -iH\tau/\hbar$ we write
\begin{equation}
  \label{glg2a}
       \langle z_{j+1} |  \re^{-\frac{\ri }{\hbar}
                   \hat{H}(t_j)\tau} | z_j \rangle
                 = \exp{ \left\{ \frac{1}{2}(
                  z_{j+1}^\star - z_j^\star )z_j - \frac{1}{2}
                  z_{j+1}^\star (z_{j+1} - z_j)-\frac{\ri \tau }
                 {\hbar}{\cal H}_{j+1,j} \right\}}
\end{equation}
where
\begin{equation}
  \label{glg40}
  {\cal H}_{j+1,j} \equiv \frac{\langle z_{j+1}|\hat{H}(t_j)|z_j\rangle }
                   {\langle z_{j+1}|z_j\rangle }
                   \equiv {\cal H}(z_{j+1}^\star,z_j;t_j)
\end{equation}
and $(1 -i {\cal H}_{j+1,j}\tau/\hbar)$ has been approximated again by
$e^{-i {\cal H}_{j+1,j}\tau/\hbar}$.  With these manipulations the
first form of the propagator, that we shall call $K_1$, becomes
\begin{align}
  \label{glg82a}
  K_1(z'',t;z',0) = \int \Bigl\{ \prod_{j=1}^{N-1}\frac{\rd^2z_j}{\pi}
                   \Bigr\}
                  \exp{ \left\{ \sum_{j=0}^{N-1} \left[ \frac{1}{2}(
                  z_{j+1}^\star - z_j^\star )z_j - \frac{1}{2}
                  z_{j+1}^\star (z_{j+1} - z_j)-\frac{\ri \tau }
                 {\hbar}{\cal H}_{j+1,j}\right] \right\}}
\end{align}
When the limit $N \rightarrow \infty$ (respectively $\tau
\rightarrow 0$) is taken, the above summations turn into
integrals, and expressions \refg{glg82a}  appears to be exact,
were it not for the well--known problems attached to the meaning
of such functional integrals. Also, ${\cal H}_{j+1,j}$ turns into
the smooth Hamiltonian function ${\cal H}(z,z^\star) \equiv
H_1(z,z^\star) \equiv \langle z |\hat{H}|z\rangle$. Using the
properties $\hat{a}|z\rangle = z|z\rangle$ and $\langle
z|\hat{a}^\dagger  = \langle z | z^\star\,$, we see that ${\cal
H}$ can be easily calculated if $\hat{H}$ is written in terms of
creation and annihilation operators with all $\hat{a}^\dagger\,$'s
to the left of the $\hat{a}\,$'s. Therefore, ${\cal H}$ is exactly
the Q symbol of the Hamiltonian operator \cite{Hill84}.

The second form of path integral starts from the ``diagonal
representation'' of the hamiltonian operator, namely
\begin{equation}
  \label{5mb1}
\hat{H} = \int|z\rangle H_2(z^\star,z)\frac{\rd^2z}{\pi}\langle z| \; .
\end{equation}
Assuming that $\hat{H}$ is either a polynomial in $p$ and $q$ or a
converging sequence of such polynomials, this diagonal
representation always exists. The calculation of $H_2$ is not as
direct as that of $H_1$, but it can be shown \cite{Hill84} that
$H_2(z^\star,z)$ is exactly the P symbol of $\hat{H}$. This second
form will be contrasted with the first--form hamiltonian function
$H_1(z^\star,z)$. To facilitate the comparison between the second
form of the propagator, that we call $K_2$, and the first form
$K_1$, it is convenient to break the time interval $T$ into $N-1$
intervals, rather than $N$. We write
\begin{eqnarray}
 \label{form2a}
K_2(z'',T;z',0)  = \langle z'' | \prod_{j=1}^{N-1}
\re^{-\frac{i\tau}{\hbar}\hat{H}}\,|  z' \rangle
\end{eqnarray}
and, following Klauder and Skagerstam, we write the infinitesimal
propagators as
\begin{equation}
  \label{5mb7}
\re^{-\frac{\ri}{\hbar}\hat{H}\tau} \approx
\int|z_j\rangle
  \left( 1-\frac{i\tau}{\hbar}H_2(z_j^\star,z_j) \right)
   \frac{\rd^2z_j}{\pi}\langle z_j|
\approx \int|z_j\rangle
   \re^{-\frac{i\tau}{\hbar}H_2(z_j^\star,z_j)}
   \frac{\rd^2z_j}{\pi}\langle z_j| \,.
\end{equation}
The complete propagator $K_2$ becomes
\begin{align}
  \label{5mb8a}
K_2(z_N,&T;z_0,0) =  \displaystyle{
    \int  \prod_{j=1}^{N-1}\frac{\rd^2z_j}{\pi}
     \langle z_{j+1} | z_j \rangle \,
     \exp{ \left\{ -\frac{\ri \tau } {\hbar} H_2(z_j^\star,z_j) \right\}}}
     \nonumber \\
 &=  \int  \{ \prod_{j=1}^{N-1}  \frac{\rd^2z_j}{\pi} \}
             \exp{ \left\{  \sum_{j=0}^{N-1} \left[ \frac{1}{2}(
                  z_{j+1}^\star - z_j^\star )z_j - \frac{1}{2}
                  z_{j+1}^\star (z_{j+1} - z_j)
             -\frac{\ri \tau } {\hbar} H_2(z_j^\star,z_j) \right]
              \right\}}  \; .
\end{align}

The differences between $K_1$ and $K_2$ are subtle but important.
While the two arguments of $H_1$ in $K_1$ belong to two adjacent
times in the mesh, the two arguments of $H_2$ in $K_2$ belong to
the same time.  Although both forms should give identical results
when computed exactly, the differences between the two are
important for the stationary exponent approximation, resulting in
different semiclassical propagators. The semiclassical evaluation
of $K_1$ and $K_2$ were presented in detail in \cite{Bar01}. Here
we only list the results:
\begin{eqnarray}
  \label{glg119}
  K_1(z'',t;z',0) = \sum_\nu
         \sqrt{\frac{\ri}{\hbar}\frac{\partial^2 S_{1\nu}}
         {\partial u' \partial v''}} \;
         \exp \left\{ \frac{\ri}{\hbar}(S_{1\nu}+I_{1\nu}) - \frac{1}{2}
          \bigl( |z''|^2 + |z'|^2\bigr) \right\} \, ,
\end{eqnarray}
\begin{eqnarray}
  \label{glg119a}
  K_2(z'',t;z',0) = \sum_\nu
         \sqrt{\frac{\ri}{\hbar}\frac{\partial^2 S_{2\nu}}
         {\partial  u' \partial v''}} \;
         \exp \left\{ \frac{\ri}{\hbar}(S_{2\nu}-I_{2\nu}) - \frac{1}{2}
          \bigl( |z''|^2 + |z'|^2\bigr) \right\} \, ,
\end{eqnarray}
where
\begin{eqnarray}
  \label{glg83}
  S_{i\nu} = S_{i\nu}(v'',u',t)
         &=\int\limits_0^t \rd t' \left[\frac{\ri \hbar}{2}
          (\dot{u}v-\dot{v}u) - H_i(u,v,t') \right]
          - \frac{\ri \hbar}{2} ( u''v'' + u'v' )
\end{eqnarray}
is the action and
\begin{equation}
\label{epslb}
I_i = \frac{1}{2} \int_0^T
\frac{\partial^2 H_i}{\partial u \partial v}
{\rm d}t
\end{equation}
is a correction to the action. The sum over $\nu$ represents the sum
over all (complex) classical trajectories satisfying Hamilton's
equations
\begin{eqnarray}
  \label{glg10}
 \ri \hbar\dot{u}&= + \displaystyle{\frac{\partial H_i}{\partial v}}\nonumber\\
 \ri \hbar\dot{v}&= - \displaystyle{\frac{\partial H_i}{\partial u}}
\end{eqnarray}
with boundary conditions
\begin{equation}
\label{mb12} u(0) = z'\equiv u'~, \qquad v(t) = {z''}^\star
\equiv v"~ \;.
\end{equation}
The factors $I_i$ are an important part of the above formulas and
they are absolutely necessary to recover the exact propagator for
quadratic Hamiltonians. If one neglects it, even the Harmonic
oscillator comes out wrong. For a discussion about
non-contributing trajectories, see refs. \cite{Rib04,Ada89}.

%%%%%%%%%%%%%%%%%%%%%%%%%%%%%%%%%%%%%%%%%%%%%%%%%%%%%%%%%%%%%%%%%%%
\subsection{The Conjectured Weyl approximation}

As discussed at the begining of this section, for a given quantum
operator $\hat{H}$, the first-form Hamiltonian is given by
$H_1(z^\star,z)= \langle z|\hat{H}|z\rangle$. It can be obtained
by writing $\hat{H}$ in terms of the operators $\hat{a}$ and
$\hat{a}^\dagger$ in {\em normal order}, so that each monomial in
$\hat{H}$ look like $c_{nm} \hat{a}^{\dagger n}\hat{a}^m$.
Replacing $\hat{a}^\dagger$ by $z^\star$ and $\hat{a}$ by $z$
yields $H_1$. $H_2$ can be calculated by writing $\hat{H}$ in {\em
anti-normal order}, where now each monomial looks like $c_{nm}
\hat{a}^n \hat{a}^{\dagger m}$, and then replacing
$\hat{a}^\dagger$ by $z^\star$ and $\hat{a}$ by $z$. A third type
of Hamiltonian function can be obtained from $\hat{H}$ by using
the Wigner transformation:
\begin{equation}
  \label{5mb18}
H_W(q,p) = \int\rd s\,\re^{\frac{\ri}{\hbar}ps}\left\langle q-\frac{s}{2}\left|
   \hat{H}\right|q+\frac{s}{2}\right\rangle \; .
\end{equation}
This is the Weyl Hamiltonian. Since $H_W$ is obtained from
$\hat{H}$ by completely symmetrizing the creation and annihilation
operators, it turns out that $H_W$ is an exact average between
$H_1$ and $H_2$ {\it if} $\hat{H}$ contains up to cubic monomials
in $\hat{a}$ and $\hat{a}^\dagger$, but only an approximate
average for other cases. The semiclassical formula with $H_1$
comes with a correction $+I_1$ to the action and that with $H_2$
comes with a correction of $-I_2$. This suggests a third type of
semiclassical approximation for the propagator, where one uses the
Weyl Hamiltonian and no correction term, since the average of
$+I_1$ and $-I_2$ should be approximately zero. This is the Weyl
approximation, which was conjecture in \cite{Bar01}:
\begin{eqnarray}
  \label{glg119b}
  K_W(z'',t;z',0) = \sum_\nu
         \sqrt{\frac{\ri}{\hbar}\frac{\partial^2 S_{W}}
         {\partial u' \partial v''}} \;
         \exp \left\{ \frac{\ri}{\hbar}S_{W} - \frac{1}{2}
          \bigl( |z''|^2 + |z'|^2\bigr) \right\} \,
\end{eqnarray}
with $S_W$ given by Eq.(\ref{glg83}) with $H_i$ replaced by $H_W$.

Of the three semiclassical approximations presented, the Weyl
approximation seems to be the most natural, since it involves the
classical hamiltonian directly and no corrections to the action.
However, this formula does not follow from the two most natural
forms of path integral proposed by Klauder and used in this
section. In the next section we propose a third form of path
integral whose semiclassical limit is indeed the Weyl
approximation.

%%%%%%%%%%%%%%%%%%%%%%%%%%%%%%%%%%%%%%%%%%%%%%%%%%%%%%%%
%%%%%%%%%%%%%%%%%%%%%%%%%%%%%%%%%%%%%%%%%%%%%%%%%%%%%%%%
\section{A Mixed Form for the Path Integral}

The new form of path integral we describe in this section is based
on the fact that $H_W$ is almost the average of $H_1$ and $H_2$.
The idea is to force this average to appear by combining the first
and second form of path integrals in alternating time steps. We
start from
\begin{equation}
  K(z'',t;z',0) = \langle z_N | \prod_{j=0}^{N-1}
                   \re^{-\frac{\ri }{\hbar}
                   \hat{H}\tau_j} | z_0 \rangle
\end{equation}
where $z_N=z"$, $z_0=z'$, $\tau_j$ is the time step and we take
$N$ to be even for convenience. Although we shall consider the
time steps $\tau_j$ to be all equal later, we keep the index $j$
for now to keep track of the time intervals.

For j odd we approximate
\begin{equation}
  \label{5mb7a}
\re^{-\frac{\ri}{\hbar}\hat{H}\tau_j} \approx \int|z_j\rangle
  \left( 1-\frac{i\tau_j}{\hbar}H_2(z_j^\star,z_j) \right)
   \frac{\rd^2z_j}{\pi}\langle z_j|
\approx \int|z_j\rangle
   \re^{-\frac{i\tau_j}{\hbar}H_2(z_j^\star,z_j)}
   \frac{\rd^2z_j}{\pi}\langle z_j| \;.
\end{equation}
For $j>0$ even we simply insert a unit operator on the right of
the infinitesimal propagator:
\begin{equation}
  \label{glg21a}
 \re^{-\frac{\ri}{\hbar}\hat{H}\tau_j}  =
   \int \re^{-\frac{\ri}{\hbar}\hat{H}\tau_j}
   |z_j\rangle\frac{\rm{d}^2z_j}{\pi}\langle z_j| \;.
\end{equation}
Multiplying this operator on the left by the bra $\langle
z_{j-1}|$ coming from the odd term $j-1$ and using the
approximation employed in the first form of path integrals,
Eq.(\ref{glg2a}), we get the following mixed form for the
propagator:
\begin {align}
  \label{5mb8}
K(z_N,T;z_0,0) &=   \int  \left\{ \prod_{j=1}^{N-1}
                \frac{\rd^2z_j}{\pi} \right\}
             \exp \left\{  \sum_{j=0}^{N-1} \left[ \frac{1}{2}(
                  z_{j+1}^\star - z_j^\star )z_j - \frac{1}{2}
                  z_{j+1}^\star (z_{j+1} - z_j) \right. \right. \nonumber \\
 &   \left. \left.  -\frac{\ri \tau } {\hbar} a_j H_{2,j}
             -\frac{\ri \tau} {\hbar}
             b_j H_{1,j} \right]
              \right\}  \nonumber \\
 & \equiv  \int \Bigl\{ \prod_{j=1}^{N-1}\frac{\rd z^\star _j\rd
                   z_j}{2\pi\ri }\Bigr\} \re^{f(z^\star ,z)}
\end {align}
where $a_j$ is zero for $j$ even and one for $j$ odd, $b_j$ is
zero for $j$ odd and one for $j$ is even. At this point we have
suppressed the index on the time intervals and have taken
$\tau_j=\tau$. The exponent $f$ is given by
\begin{align}
  \label{glg2}
   f(z^\star ,z)
          &= \sum_{j=0}^{N-1} \left\{ \frac{1}{2}(
           z_{j+1}^\star - z_j^\star )z_j - \frac{1}{2}
           z_{j+1}^\star (z_{j+1} - z_j)-\frac{\ri \tau } {\hbar}
           a_j H_{2,j}-\frac{\ri \tau }
          {\hbar}b_j H_{1,j}\right\}
\end{align}
where we have introduced the abbreviated notation $H_{2,j}\equiv
H_2(z_j^\star,z_j)$ and $H_{1,j} \equiv {\cal H}(z^*_{j+1},z_j)$.

\subsection{The Stationary Exponent Approximation}
  \label{kap2_2}

In the semiclassical limit $\hbar \rightarrow 0$ we can
approximate the integrals (\ref{5mb8}) by looking for the places
where the exponent $f$ is stationary and replacing it in their
vicinity by a quadratic form of its variables $(z^\star ,z)$. We
find the stationary points by requiring the vanishing of the
derivatives of $f$ with respect to $z$ and $z^\star$ separately.
We obtain
\begin{alignat}{3}
  \label{glg3}
    \frac{\partial f}{\partial z_j}
       &=z_{j+1}^\star - z_j^\star- \frac{\ri a_j\tau}{\hbar}\frac{\partial
        H_{2,j}}{\partial z_j} - \frac{\ri b_j\tau}{\hbar}\frac{\partial
        H_{1,j}}{\partial z_j}
       &=0 \; ; \qquad
       &j = 1,\ldots,N-1 \nonumber\\
    \frac{\partial f}{\partial z_{j+1}^\star}
       &= - z_{j+1} + z_j - \frac{\ri a_{j+1} \tau}{\hbar}\frac{\partial
        H_{2,j}}{\partial z_{j+1}^\star}- \frac{\ri b_j \tau}{\hbar}\frac{\partial
        H_{1,j}}{\partial z_{j+1}^\star}
       &=0 \; ;  \qquad
       &j = 0,\ldots,N-2 \; .
\end{alignat}

We now introduce new integration variables $\eta$ and
$\eta^\star$, which describe the deviations from the points of
stationary exponent, $z \rightarrow z + \eta\;,\ z^\star
\rightarrow z^\star + \eta^\star$, with the boundary conditions
\begin{equation}
  \label{glg41}
  \eta_0=\eta_0^\star=\eta_N=\eta_N^\star=0 \; .
\end{equation}
Expanding the exponent into a Taylor series in $(\eta^\star
,\eta)$ around the stationary points $(z^\star,z)$ up to second
order and re-inserting the result into \refg{glg82} yields
\begin{multline}
  \label{glg6}
  K(z'',t;z',0) = \re^{f(z^\star,z)}
         \mbox{\large $\displaystyle \int$} \left\{ \prod_{j=1}^{N-1}
         \frac{\rd\eta_j^\star \rd\eta_j}{2\pi \ri }
     \right\}{\rm exp} \sum_{j=0}^{N-1}\Bigl\{ -\frac{\ri\tau}{2\hbar}\
         [b_j\frac{\partial^2 H_{1,j}}
         {\partial z_j^2}+a_j\frac{\partial^2 H_{2,j}}
         {\partial z_j^2}]\eta_j^2 \\
         -\frac{\ri}{2\hbar}\,[b_j\tau \frac{\partial^2
         H_{1,j}}{\partial z_{j+1}^{\star 2}}+a_{j+1}\tau
         \frac{\partial^2
         H_{2,j+1}}{\partial z_{j+1}^{\star 2}}]\eta_{j+1}^{\star 2}\\
     -\left(1+ \frac{\ri \tau a_j }{\hbar}
         \frac{\partial^2 H_{2,j}}{\partial z_j^\star
         \partial z_j} \right)\eta_j^\star \eta_j+ \left(1- \frac{\ri \tau b_j }{\hbar}
         \frac{\partial^2 H_{1,j}}{\partial z_{j+1}^\star
         \partial z_j} \right) \eta_{j+1}^\star\eta_j \Bigr\} .
\end{multline}

The integrals in Eq.(\ref{glg6}) can be carried out using the same
techniques presented in \cite{Bar01}. The idea is to integrate
first over $\eta_1^\star$ and $\eta_1$, then over $\eta_2^\star$
and $\eta_2$, etc. A recursion formula can be readily established
and, once all integrations are done, we obtain
\begin{equation}
  \label{glg43}
  K(z_N,t;z_0,0) = \re^{f(z^\star,z)}\prod_{j=1}^{N-1} \frac{1}
         {\sqrt{\displaystyle (1+\frac{\ri \tau a_j}{\hbar}
         \frac{\partial^2H_{2,j}}
         {\partial z_j^\star \partial z_j })^2+2\ri \frac{\tau}{\hbar}
         (a_j\frac{\partial ^2H_{2,j}}{\partial z_j^2}
         +b_j\frac{\partial ^2H_{1,j}}{\partial z_j^2})X_j}}
\end{equation}
where $X_j$ satisfies
\begin{align}
  \label{glg7}
  X_j & = -\frac{\ri \tau a_j }{2\hbar}\frac{\partial^2 H_{2,j}}
         {\partial z_j^{\star 2}}
         -\frac{\ri \tau b_{j-1} }{2\hbar}\frac{\partial^2 H_{1,j-1}}
         {\partial z_j^{\star 2}}\\
        & + \frac{\displaystyle \left(
         1-\frac{\ri \tau b_{j-1}}{\hbar} \frac{\partial^2 H_{1,j-1}}
         {\partial z_j^\star \partial z_{j-1} }\right)^2}
         {\displaystyle (1+\frac{\ri \tau a_{j-1}}{\hbar}
         \frac{\partial^2 H_{2,j-1}}
         {\partial z_{j-1}^\star \partial z_{j-1} })^2 + 2\frac{\ri \tau}{\hbar}
         (a_{j-1}\frac{\partial^2 H_{2,j-1}}{\partial z_{j-1}^2}
         +b_{j-1}\frac{\partial^2 H_{1,j-1}}{\partial z_{j-1}^2})
         X_{j-1}}X_{j-1}
\end{align}
for $j=1,\ldots,N-1$ with $X_0=0$.

%%%%%%%%%%%%%%%%%%%%%%%%%%%%%%%%%%%%%%%%%%%%%%%%%%%%%%%%%%%%%%%%%%%%%%%%%%%
\subsection{The Effective Hamiltonian}
  \label{kap2.2}

To obtain the continuum limit of the discrete equations of motion
(\ref{glg3}) we first note that, since $a_j$ is zero for $j$ even
and one for $j$ odd, and $b_j$ is zero for $j$ odd and one for $j$
is even, the second of Eqs.(\ref{glg3}) gives
\begin{alignat}{3}
  \label{glg3a}
    - z_{j+1} + z_j =0
\end{alignat}
for $j$ even.

This result motivates the choice of a new time step $\epsilon =
2\tau$, in such a way that the discretized time evolution goes
from $z_j$ directly to $z_{j+2}$ for $j$ even and from $z_j^\star$
directly to $z_{j+2}^\star$ for $j$ odd . This choice of time step
makes sense if one can find an effective Hamiltonian able to
perform the corresponding evolution for both variables $z$ and
$z^*$. Indeed, the equations of motion can be put the form
\begin{alignat}{3}
  \label{glg3b}
    \frac{z_{j+1}-z_{j-1}}{\epsilon}
       &=- \frac{\ri }{\hbar}\frac{\partial
        H_{ef,j}}{\partial z_j^{\star}}
       \qquad
       &j = 1,3,5\ldots,N-3 \nonumber\\
    \frac{z_{j+1}^\star-z_{j-1}^\star}{\epsilon}
       &= \frac{\ri }{\hbar}\frac{\partial
        H_{ef,j}}{\partial z_j}
        \qquad
       &j = 2,4,\ldots,N-2 \; .
\end{alignat}
where
\begin{equation}
  \label{glg4}
  \begin{array}{ll}
    H_{ef,j} &\equiv \displaystyle{ \left(b_j H_{1,j} + b_{j-1} H_{1,j-1}
    +a_{j-1} H_{2,j} + a_j H_{2,j+1}\right)/2} \\ \\
     & = \left\{ \begin{array}{l} \displaystyle{
     \frac{H_{1,j} + H_{2,j}}{2}} \quad \mbox{for $j$ even} \\
     \displaystyle{ \frac{H_{1,j-1} + H_{2,j+1}}{2}} \quad \mbox{for $j$ odd} \;.
     \end{array} \right. \end{array}
\end{equation}
Note that, because we are skipping points with the new time step
$\epsilon=2\tau$, we miss the point $z_N^\star$ in
Eq.(\ref{glg3b}). This, however, does not affect the limit of the
continuum, as long as we take $z_{N-1}^\star = z_N^\star$ to
ensure the proper boundary condition.

In the limit where $2\tau=\epsilon$ goes to zero the effective
Hamiltonian reduces to
\begin{alignat}{3}
  \label{coe3}
    {\cal H_{C}}(z^\star,z) \equiv
      \frac{H_1(z^\star,z)+H_2(z^\star,z)}{2} \;.
\end{alignat}

As in the case of the semiclassical formulas Eqs.(\ref{glg119})
and (\ref{glg119a}) for $K_1$ and $K_2$, the stationary trajectory
is usually complex. It is therefore convenient to follow the
notation introduced in \cite{Bar01} and make the substitutions
\begin{align}
  \label{glg9}
z &\rightarrow  u = \frac{1}{\sqrt{2}}\left(\frac{q}{b}
       +\ri \,\frac{p}{c} \right) \nonumber\\
z^\star &\rightarrow  v = \frac{1}{\sqrt{2}}\left(\frac{q}{b}
       -\ri \,\frac{p}{c} \right).
\end{align}
In terms of $u$ and $v$, the stationary phase conditions
(\ref{glg3}) turn into Hamilton´s equations with ${\cal H_{C}}$:
\begin{align}
  \label{glg10a}
  \ri \hbar\dot{u} &= + \frac{\partial {\cal H_{C}}}{\partial v}\nonumber\\
  \ri \hbar\dot{v} &= - \frac{\partial {\cal H_{C}}}{\partial u}
\end{align}
with boundary conditions identical to Eq.(\ref{mb12}). The
function $f$ can also be simplified to
\begin{equation}
\label{glg15} f = \int_0^t \rd t' \left[ \frac{1}{2} (\dot{v} u -
\dot{u} v)
            - \frac{\ri}{\hbar} {\cal H_{C}}(u,v,t') \right]
+ \frac{1}{2} (v''u'' + v'u') - \frac{1}{2} ({|z''|}^2 + {|z'|}^2)
\end{equation}
where $u(0)=u'=z'$, $v(0)=v'$, $u(T)=u"$ and $v(T)=v"=z"^\star$.

Next we calculate the product appearing  in \refg{glg43}.
Performing the limit $N \rightarrow \infty$ and using the
expansion $\ln (1+x) = x + O(x^2)$ we obtain
\begin{align}
  \label{new1}
  \Gamma \equiv &\lim_{N\rightarrow\infty} \prod_{j=1}^{N-1}
  \left\{ (1+\frac{\ri \tau a_j}{\hbar}
         \frac{\partial^2H_{2,j}}
         {\partial z_j^\star \partial z_j })^2+2\ri \frac{\tau}{\hbar}
         (a_j\frac{\partial ^2H_{2,j}}{\partial z_j^2}
         +b_j\frac{\partial ^2H_{1,j}}{\partial z_j^2})X_j
          \right\}^{-\frac{1}{2}}\nonumber\\
      &=\lim_{N\rightarrow\infty} \exp\biggl\{ - \frac{1}{2}\sum_{j=1}^{N-1}
          \ln [ 1+\frac{2\ri \tau a_j}{\hbar}
         \frac{\partial^2H_{2,j}}
         {\partial z_j^\star \partial z_j }+2\ri \frac{\tau}{\hbar}
         (a_j\frac{\partial ^2H_{2,j}}{\partial z_j^2}
         +b_j\frac{\partial ^2H_{1,j}}{\partial z_j^2})X_j+O(\tau^2) ]\biggr\}\nonumber\\
      &=\lim_{N\rightarrow\infty} \exp\biggl\{ - \frac{\ri}{\hbar}\sum_{j=1}^{N-1}
          \tau[ a_j
         \frac{\partial^2H_{2,j}}
         {\partial z_j^\star \partial z_j }+
         (a_j\frac{\partial ^2H_{2,j}}{\partial z_j^2}
         +b_j\frac{\partial ^2H_{1,j}}{\partial z_j^2})X_j ]\biggr\} \; .
\end{align}
In order to transform these sums into integrals we note that
\begin{align}
  \label{new3}
  \lim_{N\rightarrow\infty}\biggl\{
       \sum_{j=1}^{N-1}\tau(a_j F(t_j))
          \biggr\} &= \lim_{N\rightarrow\infty} \tau
          (F(t_1)+F(t_3)+F(t_5)+...) \nonumber \\
          &=\frac{1}{2}\lim_{N\rightarrow\infty} \tau
          (F(t_1)+F(t_1)+F(t_3)+F(t_3)+F(t_5)+...) \nonumber \\
          &=\frac{1}{2}\lim_{N\rightarrow\infty} \tau
          (F(t_1)+F(t_2)+F(t_3)+F(t_4)+F(t_5)+...) \nonumber \\
          &= \frac{1}{2}\int\limits_0^t \rd t' \, F(t')
\end{align}
since, for smooth functions, $F(t_j)\rightarrow F(t_{j+1})$ as
$\tau \rightarrow 0$. The integrals with the coefficients $b_j$
also acquire the $1/2$ factor. Using these results we obtain
\begin{align}
  \label{new4}
  \Gamma=
  &= \exp \biggl\{ - \frac{\ri }{\hbar} \int\limits_0^t \rd t'
          \, (\frac{1}{2}\frac{\partial^2H_2}{\partial
          u\partial v}(t')+\frac{1}{2}(\frac{\partial^2 H_1}{\partial
          u^2}+\frac{\partial^2 H_2}{\partial
          u^2}) X(t') )\biggr\} \nonumber\\
  &= \exp \biggl\{ - \frac{\ri }{\hbar} \int\limits_0^t \rd t'
          \, [\frac{1}{2}\frac{\partial^2H_2}{\partial
          u\partial v}(t')+\frac{\partial^2 {\cal H_{C}}}
          {\partial u^2}(t') X(t') ]\biggr\} \; .
\end{align}

%%%%%%%%%%%%%%%%%%%%%%%%%%%%%%%%%%%%%%%%%%%%%%%%%%%%%%%%%%%%%%%%%%%%%%%%%%%
\subsection{The Effective Phase}
  \label{kap2.3}

Replacing $\Gamma$ and $f$ into Eq.(\ref{glg43}) we obtain
\begin{align}
  \label{glg18}
  K(z'',t;z',0)
    &= \exp \biggl\{ - \frac{\ri }{\hbar} \int\limits_0^t \rd t'
          \, [\frac{1}{2}\frac{\partial^2{\cal H}_{2}}{\partial
          u\partial v}(t')+\frac{\partial^2 {\cal H_{C}}}
          {\partial u^2}(t') X(t') ]\biggr\} \nonumber\\
    & \exp \biggl\{ \int\limits_0^t \rd t'
          \left[ \frac{1}{2}\left(\dot{v}u -
          \dot{u}v\right) -\frac{\ri }{\hbar}{\cal H_{C}} \right]
           + \frac{1}{2}\left(v'u' + v''u''\right) -
          \frac{1}{2}\left(|z'|^2 + |z''|^2\right) \biggr\} \; .
\end{align}

We still have to write the continuous form of the discrete
recursion formula (\ref{glg7}) for $X(t)$. In the limit $N
\rightarrow \infty$ we obtain the nonlinear differential equation
\begin{align}
  \label{glg17}
  \dot{X}(t) = -\frac{\ri }{2\hbar}\frac{\partial^2{\cal H_{C}}}{\partial
          v^2}-2\frac{\ri }{\hbar}\frac{\partial^2{\cal H_{C}}}{\partial
          u\partial v}X(t) -2\frac{\ri }{\hbar} \frac{\partial^2{\cal H_{C}}}
          {\partial u^2} X^2(t)
\end{align}
with the initial condition $X(0)= 0$.

This equation was solved in \cite{Bar01} and the result is $X =
\frac{1}{2} \frac{\delta u}{\delta v}$ where $\delta u$ and
$\delta v$ are solutions of the linearized Hamilton's equations
\begin{align}
  \label{glg30}
  \delta \dot{u} &= -\frac\ri {\hbar}\frac{\partial^2{\cal H_{C}}}{\partial u
          \partial v} \,\delta u - \frac\ri {\hbar}\frac{\partial^2{\cal H_{C}}}
          {\partial v^2}\,\delta v\nonumber\\
  \delta \dot{v} &= +\frac\ri {\hbar}\frac{\partial^2{\cal H_{C}}}{\partial u^2}
          \,\delta u + \frac\ri {\hbar}\frac{\partial^2{\cal H_{C}}}{\partial u
          \partial v} \,\delta v
\end{align}
where the derivatives are calculated at the stationary trajectory
and the initial conditions are $\delta u(0)=0$ and $\delta v(0)$
arbitrary. The second term in the first exponential of
\refg{glg18} can be now transformed with the help of \refg{glg30}
\begin{align}
  \label{glg32}
  \frac\ri {\hbar}\frac{\partial^2{\cal H_{C}}}{\partial u^2}X
         &= \frac\ri {2\hbar}\frac{\partial^2{\cal H_{C}}}{\partial u^2}
          \frac{\delta u}{\delta v}
          =\frac{1}{2}\frac{\delta \dot{v}}{\delta v}-\frac\ri {2\hbar}
          \frac{\partial^2{\cal H_{C}}}{\partial u\partial v}
          =\frac{1}{2}\frac{\rm{d}}{\rm{d}t}\ln \delta v - \frac\ri
          {2\hbar}\frac{\partial^2{\cal H_{C}}}{\partial u\partial v}
\end{align}
so that the first exponent of \refg{glg18} becomes
\begin{align}
  \label{glg33}
      & \exp \left\{ -\frac{\ri}{\hbar}\mbox{\large $\displaystyle
          \int$}^t_{\!\!\!\!0} \rd t' [\frac{1}{2}\frac{\partial^2H_2}
          {\partial u\partial v}
          (t' )+\frac{\partial^2{\cal H_{C}}}{\partial u^2}
          (t' )X(t' )] \right\} \\
         & = \exp \left\{ +\frac{\ri}{4\hbar}\mbox{\large $\displaystyle
          \int$}^t_{\!\!\!\!0} \rd t' [\frac{\partial^2{{\cal H}_{1}}}
          {\partial u\partial v}
          (t' )-\frac{\partial^2{{\cal H}_{2}}}
          {\partial u\partial v}
          (t' )] \right\}\exp \left\{ -\frac{1}{2}\mbox{\large $\displaystyle
          \int$}^t_{\!\!\!\!0} \rd t' \left[ \frac{\rd}{\rd t'}
          (\ln \delta v) \right]  \right\}  \nonumber  \\
         & = \sqrt{\frac{\delta v'}{\delta v''}}
         \displaystyle{\exp{\left[ \frac{\ri} {\hbar} \left(\frac{I_{1}-I_{2}}{2}
         \right)\right]}} \equiv \sqrt{\frac{\delta v'}{\delta v''}}
         \displaystyle{\exp{\left[ \frac{\ri} {\hbar} I_C \right]}} \;.
\end{align}

The pre-factor $\delta v'/\delta v''$ can also be written in terms
of the action using $\partial S_C/\partial u' = -i\hbar v'$, where
\begin{eqnarray}
  \label{coe00}
  S_{C}(v'',u',t) :
         &=\int\limits_0^t \rd t' \left[\frac{\ri \hbar}{2}
          (\dot{u}v-\dot{v}u) - H_C(u,v,t') \right]
          - \frac{\ri \hbar}{2} ( u''v'' + u'v' )
\end{eqnarray}
is the effective action. In the end we obtain
\begin{align}
  \label{new6}
  K(z'',t;z',0) &= \sqrt{\frac{\ri}{\hbar}\frac{\partial^2 S_{C}}
             {\partial u' \partial v''}}
            \exp \biggl\{\frac{\ri}{\hbar}(S_C + I_C )-
            \frac{1}{2}(|z'|^2 +|z''|^2)\biggr\}
\end{align}
where
\begin{align}
  \label{efcph}
I_C = \frac{1}{2}(I_1-I_2)
\end{align}
is the effective phase.

%%%%%%%%%%%%%%%%%%%%%%%%%%%%%%%%%%%%%%%%%%%%%%%%%%%%%%%%
%%%%%%%%%%%%%%%%%%%%%%%%%%%%%%%%%%%%%%%%%%%%%%%%%%%%%%%%
\section{The Weyl Approximation}

Equation (\ref{new6}) is the main result of this paper. It
represents a third semiclassical approximation for the coherent
states propagator, involving the effective Hamiltonian
$H_C=(H_1+H_2)/2$ and the effective phase $I_C=(I_1-I_2)/2$.

For the harmonic oscillator $I_{1}$ and $I_{2}$ are exactly equal
and $I_{C}=0$. In this case $H_C$ coincides with the classical, or
Weyl, Hamiltonian $H_W$, and the conjectured Weyl approximation
(\ref{glg119b}) is obtained. In fact $I_{1}=I_{2}$ and $H_C=H_W$
for all polynomial Hamiltonians involving up to cubic powers of
$q$ or $p$ \cite{Bar01}. This can be seen from the formulas
\cite{Bar01,Klau85}
\begin{alignat}{3}
  \label{coe10}
    H_1(z^\star,z)
    = \exp{\left({\frac{1}{2}\hat{\delta}}\right)} H_W(z^\star,z)
\end{alignat}
\begin{alignat}{3}
  \label{coe11}
    H_2(z^\star,z)
    =\exp{\left(-\frac{1}{2}\hat{\delta}\right)} H_W(z^\star,z)
\end{alignat}
where $\hat{\delta}=\partial^2 / \partial z^\star \partial z$.
This gives
\begin{alignat}{3}
  \label{new9}
    H_C
    = \cosh{\left(\frac{1}{2}\hat{\delta}\right)} H_W
    = H_W - \frac{1}{8}\frac{\partial^4}{\partial z^2
    \partial {z^\star}^2} H_W + ...
\end{alignat}
which shows explicitly that $H_C=H_W$ for up to cubic polynomials.
Besides, using the relations
\begin{equation}
  \label{coe5}
q = (z+z^\star)( \frac{b}{\sqrt{2}}) \qquad p= (z-z^\star)(
\frac{-i\hbar}{b\sqrt{2}})
\end{equation}
with $b \sim c \sim O(\hbar^{1/2})$, we see that quartic or higher
order terms contribute to $H_C$ or $I_C$ only terms of order
$\hbar^2$. These terms can in principle be neglected, since they
are beyond the scope of the approximation. With these
considerations we can rewrite the propagator (\ref{coe00}) as
\begin{align}
  \label{coe0}
  K(z'',t;z',0) &= \sqrt{\frac{\ri}{\hbar}\frac{\partial^2 S_{W}}
             {\partial u' \partial v''}}
            \exp \biggl\{\frac{\ri}{\hbar}S_W -
            \frac{1}{2}(|z'|^2 +|z''|^2)\biggr\}
\end{align}
where $S_W$ is given by Eq.~(\ref{coe00}) with $H_C$ replaced by
$H_W$. This is the Weyl formula conjectured in \cite{Bar01}.

As a final remark we notice that the differences between the three
semiclassical formulas presented in this paper are of the order of
$\hbar$, and go to zero in the semiclassical limit. These
differences, however, are always relevant at low energies, and can
be made explicit by considering the Fourier transform of these
time dependent formulas. A discussion of the energy representation
of these semiclassical propagators were presented in section 6 of
ref.\cite{Bar01}, including the derivation of semiclassical
quantization rules, and we refer to it for further details. \\ \\ \\

\noindent ACKNOWLEDGMENTS

MAMA acknowledges financial support from the Brazilian agencies
FAPESP and CNPq.

%%%%%%%%%%%%%%%%%%%%%%%%%%%%%%%%%%%%%%%%%%%%%%%%%%%%%%%%
%%%%%%%%%%%%%%%%%%%%%%%%%%%%%%%%%%%%%%%%%%%%%%%%%%%%%%%%

\end{document}